\documentclass[mathleft
]{an}
\usepackage{graphicx}
\usepackage{times}
\begin{document}

\Pagespan{284}{287}
\Yearpublication{2008}%
\Yearsubmission{2007}%
\Month{02}%
\Volume{329}%
\Issue{3}%
\DOI{10.1002/asna.200710957}%

\title{Towards a Real-time Transient Classification Engine}

\author{J. S. Bloom\inst{1}\fnmsep\thanks{Corresponding author:
  \email{jbloom@astro.berkeley.edu}\newline}
\and  D. L. Starr\inst{1,2}
\and  N. R. Butler\inst{1}
\and  P. Nugent\inst{1,3}
\and  M. Rischard\inst{1}
\and  D. Eads\inst{4}
\and D. Poznanski\inst{1}
}
\titlerunning{Time series Classification Engine}
\authorrunning{J. S. Bloom et al.}
\institute{
Astronomy Department, University of California, Berkeley, 
Berkeley, CA 94709 USA
\and 
Las Cumbres Global Telescope Network,
6740 Cortona Dr.
Santa Barbara, CA 93117 USA 
\and 
Lawrence Berkeley National Laboratory 
MS 50F-1650
1 Cyclotron Road 
Berkeley, CA, 94720 USA
\and
University of California, Santa Cruz
1156 High Street
Santa Cruz, CA 95064 USA
}

\received{2007 Oct 17}
\accepted{2008 Jan 1}
\publonline{2008 Feb 25}

\keywords{methods: statistical --- methods: data analysis --- surveys}

\abstract{%
Temporal sampling does more than add another axis to the vector of observables. Instead, under the recognition that how objects change (and move) in time speaks directly to the physics underlying astronomical phenomena, next-generation wide-field synoptic surveys are poised to revolutionize our understanding of just about anything that goes bump in the night (which is just about everything at some level). Still, even the most ambitious surveys will require targeted spectroscopic follow-up to fill in the physical details of newly discovered transients. We are now building a new system intended to ingest and classify transient phenomena in near real-time from high-throughput imaging data streams. Described herein, the {\it Transient Classification Project} at Berkeley will be making use of classification techniques operating on ``features'' extracted from time series and contextual (static) information. We also highlight the need for a community adoption of a standard representation of astronomical time series data (ie. ``VOTimeseries'').}

\maketitle

\section{Introduction}
Classification and knowledge extraction from large imaging surveys of the static sky is a maturing endeavor. Star-galaxy classification and photometric redshift estimates are well-posed problems where knowledge of the underlying physics allows for robust estimates of uncertainty.  Automated classification is critical for source demographics, especially informing large statistical questions of the data. More recently, the autonomous classification with time domain surveys have had some notable successes, particularly with microlensing surveys \cite{2003AcA....53..291U} and supernova searches \cite{2006NewAR..50..436C,2007ApJ...666..674M,2007arXiv0708.2749F}. Indeed, optimizing reductions, survey cadence strategies, and software algorithms in the search of a specific class of phenomena has been the most straightforward use of synoptic imaging (see Bailey, this workshop). While interested primarily in supernovae, the deep lens survey (\cite{2004ApJ...611..418B}; see Becker, this workshop) is one of the few large-field synoptic surveys that attempted to provide multi-class inferences in real-time. Still, those classifications were rather broadbrushed in physical scope (``supernova'', ``fast moving'', ``slow moving'', ``variable star'', ``unknown (stationary)''). With the advent of large-scale multicolor imaging surveys to appreciable depths (e.g., DES, Pan-STARRS1, Skymapper, \& LSST), the need for a real-time and general classification scheme of astronomical transient is particularly pressing.

The challenges in time-domain classification may be sub-divided into ``discovery'' and ``inference.'' Discovery of a variable or a moving source requires at least two images with the same filter system. The characteristic separation in time will dictate the types of sources that are seen to vary:  taken too close in time, moving solar system sources do not change their apparent angular position enough to recognize change; taken too distant in time, slowly moving solar system sources would be single apparition point sources, confused for extra-solar events. Without the benefit of filtering techniques with several images of a static sky, transient discovery on a few images is particular prone to cosmetic defects in the imaging arrays, cosmic rays, and near-field (non-astrophysical) interlopers\footnote{Having more than one concurrent image from at least two sites greatly reduces false positives (as employed with RAPTOR and TAOS experiments).}. Transient discovery can be performed either in ``catalog space'' (noting significant changes of source brightness in two epochs) or with source extraction in image differences \cite{2007ApJ...665.1246B}. The former is generally less computationally intensive but more susceptible to error due to variable seeing and imaging array defects. Image differencing is generally more robust in crowded fields and where transients are embedded in galaxy light.

Once a transient source is discovered, we wish to surmise the nature of the source. Major features of a general classification scheme are identified:
\begin{itemize}
	\item The inferences about the physical nature of the source (and the source variability) should make full use of prior knowledge about transients without coercing every new transient into a predefined set of classes. Different users should be allowed to tune their priors as the science requires.
	\item The inferences will necessarily be probabilistic in nature and should evolve in time as more observations are obtained. 
	\item The classifications should be as near real-time as possible to allow appropriate follow-up.
	\item The classifications should allow feedback from end users and adapt the classification algorithms accordingly.
\end{itemize}
We are now building a framework that will be capable of classifying transient sources from time-domain surveys with these features. Similar work in a machine-learning context has been reported elsewhere (e.g., \cite{2002SPIE.4846..147W,2004AN....325..477V} and Mahabal, this workshop; Bailey, this workshop). There are several components to this ``Transient Classification Project'' (TCP), described herein. We aim to have a working system in place by the time that the Palomar Transients Factory (PTF; \cite{srk07}) comes on-line in Fall 2008.

\section{Data Ingest}
The starting point of transient classification, from the perspective of the TCP, is a stream of metadata describing the individual detections (either from image differences or catalog detections). We coerce this metadata stream to an internal data model using a custom translation client written for each survey. Since we have been developing the TCP using, primarily, the public data from the SDSS-II stripe 82 survey \cite{2007AJ....134..973I}, we found it necessary to recalibrate the photometry and astrometry from the raw detection files. Objects from all surveys are ingested into a relational database with the object positions indexed using the hierarchical triangle mesh (HTM; \cite{2001misk.conf..638O}) at depth of 14 and 25 to allow for fast and accurate searching.

Object detections need to be associated with astrophysical sources. Unlike with static or after-the-fact time-domain surveys where a filtered deep sky image may be used as the fiducial ``true'' representation of source brightnesses and source positions, a real-time time-domain survey necessarily must associate each object detection with an astrophysical source. We create sources on-the-fly using a probabilistic framework that asks the question whether a new object belongs to an existing source or demands the creation of a new source.  For a new object with detected position $(O_\alpha \pm \sigma_{O,\alpha}, O_\delta \pm \sigma_{O,\delta})$ we can find a set of possible associated sources $\left\{S\right\}$ by searching in the source catalog for sources with positions with angular distance\footnote{This is quite robust if the uncertainties of the source positions in the current source catalog are similar (typically $<$1\arcsec) and comparable or less than the object positional uncertainties} $d < d_0 \sqrt{\sigma_{O,\alpha} \sigma_{O,\delta}}$ of $O$; typically we use $d_0 \approx 10$. For each source $S_i$ in $\left\{S\right\}$ we compute the logarithm of the odds ratio $\log o_i$ comparing the hypotheses that the object belongs to that source or should be a separate source. Under the assumption of Gaussianity,
\begin{eqnarray}
-2 \ln o_i &=& 
   \frac{\left(\mu_\alpha - \bar S_\alpha\right)^2}{\sigma_{S,\alpha}^2} +
                    \frac{\left(\mu_\alpha - O_\alpha\right)^2}{\sigma_{O,\alpha}^2} + \\
		 & &		\frac{\left(\mu_\delta - \bar S_\delta\right)^2}{\sigma_{S,\delta}^2} +
					\frac{\left(\mu_\delta - O_\delta\right)^2}{\sigma_{O,\delta}^2} \nonumber
\end{eqnarray}
with $\bar S_{\alpha,\delta}$ equal to the weighted mean position of $O$ and $S_i$. Since we expect $-2 \ln o_i$ to be distributed as $\chi^2$ with 2 degrees of freedom we associate the object with the $S_i$ meeting some predetermined probability threshold. As the number of objects associated with a source grows, the number of lower probability associations should also grow; through a simulation, we have found the correct number of sources are created if we change the probability threshold in accordance with the number of sources already associated with that object.

\section{Source Classification}
\subsection{Representation}
The creation or modification of a source triggers a series of steps that will lead to an updated statement about the nature of that source. In an effort to modularize the software tasks, and prepare for the possibility of distributing the computational tasks (see \ref{sec:future}), we have build a portable (XML-based) source container (which we called ``VOSource''), consisting of rudimentary source position, results from survey queries (such as NED), and the time series photometry associated with the source. The time series (which we call ``VOTimeseries'' is marked up as a VOTable\footnote{ http://www.ivoa.net/twiki/bin/view/IVOA/IvoaVOTable}, similar to the way in which time series data are represented in the VizieR\footnote{http://vizier.u-strasbg.fr/viz-bin/VizieR} catalogs. An example of a ``VOSource'' container is:
 
\begin{footnotesize}
\begin{verbatim}
<VOSOURCE version="0.01">
   <COOSYS equinox="J2000" epoch="J2000" 
          system="eq_FK5"/>
   <dbID>62</dbID>
   <WhereWhen>
      <Description>current position</Description>
      <Position2D unit='deg'>
        <!-- same as VOEvent positions -->        
      </Position2D></WhereWhen>
   <VOTIMESERIES version="0.01">
    <TIMESYS><TimeType 
      ucd="frame.time.system?">MJD</TimeType> 
     <TimeSystem ucd="frame.time.scale">
      UTC</TimeSystem> 
     <TimeRefPos ucd="pos;frame.time">
      TOPOCENTER</TimeRefPos>  </TIMESYS>
    <RESOURCE name='db photometry'>
     <TABLE name='sdss-i'>
      <FIELD name='t' ID='col1' system='TIMESYS' 
          datatype='float' unit='day'/>
      <FIELD name='m' ID='col2' 
          ucd='phot.mag;em.opt.i' 
          datatype='float' unit='mag'/>
      <FIELD name='m_err' ID='col3' 
           ucd='stat.error;phot.mag;em.opt.i' 
           datatype='float' unit='mag'/>
        <DATA><TABLEDATA>
            <TR row='1'><TD>53352.1454934000</TD>
             <TD>20.0757000000</TD>
             <TD>0.0749128000</TD></TR>
               ....
\end{verbatim}
\end{footnotesize}

\subsection{Feature Extraction: Mixing the Time-domain with Context}
The time series of a source is anything but standard --- data are irregularly sampled, noisy, sometimes spurious, and may include detections as well as non-detections. In light of this, we seek to homogenize the time series data by extracting ``features.'' A feature is a real-number line mapping, often involving basic statistical metrics on the time series (such as $\chi^2$ per degree of freedom or skewness). We are developing a custom {\it Python} codebase which is capable of ingesting a VOTimeseries and returning the features of that time series. For sources with too few observations appropriate for a given feature (such as ``largest significant peak in a Lomb-Scarle periodogram''), the results from that specific feature extraction is reported as undefined.

One of the benefits of mapping heterogeneous information to a series of real-number lines is that the nature of that information is abstracted from operations performed on those feature vectors farther downstream. To this end, (static) context related to a transient source (e.g., the location with respect to the supergalactic plane, distance to nearest cataloged galaxy, redshift of that galaxy, etc.) can play a powerful discriminating role. For example, a new point source which is discovered close to the ecliptic plane and a region of significant Galactic extinction is much more likely to be a slow moving solar system object than a distant supernova.

\subsection{Rapid Identification and Adaptation Using Machine Learning}
The TCP will make use of prior knowledge of time series and contextual information for each known class of transient. To do so, we are assembling a large labeled training set of real-world examples of known classes. To create set of prior feature distributions, these sources will be degraded and sampled with cadences and sensitivities typical of the survey. A new source and its subsequently derived feature vector can be compared directly to the priors. We do not yet know what family of classifiers will prove the most robust (see Mahabal, this workshop, for an extensive discussion of current techniques). However, we are exploring the use of pairwise Naive Bayesian voting techniques as a fast approach capable of yielding probabilistic statements about the nature of new transients. Online learning algorithms \cite{723908} (such as ``shifting experts'' \cite{296382}) and ensemble algorithms (such as ``boosting'' \cite{337870}) should be particularly applicable, since these classes of algorithms allow quick updates of the classification inferences without needing to re-analyze all the available data. When a new event arrives that cannot be characterized by existing classes, methods for identifying the anomaly and incorporating it into a new class are being considered.

\section{Future Steps}
\label{sec:future}
The TCP is a work in progress but the basic architectural decisions have now been put in place. Aside from the development and testing of the machine learning techniques, there are several other 
elements we hope the implement in the upcoming year:
\begin{itemize}
	\item {\bf Multi-survey Footprint Server}. Non-detections of a transient source provide valuable constraints for classification.  A footprint server, providing upper-limits for a given position, time and filter, will be therefore crucial to the classification algorithms
	\item {\bf Distribution}. We plan to make full use of the VOEventNet architecture to distribute newly-classified transients to TCP clients (using a variety of push/pull mechanisms)
	\item {\bf Feedback Mechanisms} We will require a formalized conduit for end users (on the receiving end of probabilistic classifications) to feed back  in to the system the outcome of transient followup.
	\item {\bf Massively Distributed Computing} We are scoping the use of the BOINC architecture\footnote{http://boinc.berkeley.edu/} to create a ``TCP@Home'' environment, where individual users will provide spare CPU cycles to crunch feature extraction methods and run classification algorithms.
\end{itemize}

While especially suited for the PTF, this classification engine is being built to not only allow several surveys streams to flow through the system but allow the information extracted in each stream to inform the classifications derived from other surveys. Since the implementation of the feature extraction and classification algorithms is atomized, we expect TCP to scale well to the data rates advertised for LSST.

\acknowledgements
We thank Las Cumbres Global Telescope Network for partial material support of this effort. JSB is grateful for the receipt of the Hellman Family Fund and Sloan Research grant, which will provide additional support for this endeavor. DP is supported by a SciDAC grant from the Department of Energy. NB and PN are partially supported by this SciDAC grant. The initial stages of this project were borne out of conversations and support from the VOEventNet collaboration. Conversations with A.~Mahabal and R.~Williams have been particularly fruitful.


\begin{scriptsize}

\end{scriptsize}

\end{document}